\documentclass{ws-p8-50x6-00}
\usepackage[T1]{fontenc}   
\usepackage{amsmath}
\usepackage[wspc]{refstyle}
\usepackage{latexuseful2e}
\usepackage{graphicx}

\newcommand{\nuNA}{\nu_{{}_{N_A}}}
\newcommand{\SNA}{S_{{}_{N_A}}}
\newcommand{\Jpsi}{J/\psi}                
\newcommand{\Npart}{N_{\text{part}}}    

\begin{document}

\title{Particle rapidity density and collective phenomena in 
       heavy ion collisions}

\author{\underline{R. Ugoccioni}\footnotemark ~~AND J. Dias de Deus}

\address{CENTRA and Departamento de F\'{\i}sica (I.S.T.) \\
Av. Rovisco Pais, 1049-001 Lisboa, Portugal}

\maketitle

\abstracts{
We analyse recent results on charged particle pseudo-rapidity
densities from RHIC in the framework of the Dual String Model, in
particular when including string fusion.
The model, in a simple way, agrees with all the existing data and is
consistent with the presence of the percolation transition to the
Quark-Gluon Plasma already at the CERN-SPS.
It leads to strict saturation of the particle
(pseudo-)rapidity density, normalised to the number of participant
nucleons, as that number increases. Asymptotically, as $\sqrt{s} \to
\infty$, with the number of participants fixed, this density
approaches again nucleon-nucleon density.
A comparison with recent WA98 data is presented.
}

The dependence of measurable quantities like
charged particle density, transverse energy and $\Jpsi$ production rate
on the number $\Npart$ of participant nucleons in high energy heavy
ion collisions is
extremely important both for a better understanding of the
initial conditions in the evolution of newly created dense matter and
because it provides the information for discriminating among 
different models.\cite{here:Gyulassy,phobos:1,WA98:1,Lourenco,Wang,RU:dNdeta}
In this contribution we analyse such quantities in the framework
of the Dual String Model (DSM).
\footnotetext{* \itshape Present address: 
Dipartimento di Fisica Teorica,
via Giuria 1, 10125 Torino, Italy}

We start by building
nucleus-nucleus collisions as resulting from superposition of
nucleon-nucleon collisions, in the way it is done in the
Glauber model approach and its generalisations:
in the DSM, i.e., the Dual Parton Model\cite{DPM:1}
with the inclusion of strings,\cite{StringFusionModel}
the valence quarks of the nucleon produce particles, via strings, only
once ---this is the wounded nucleon model case--- and production is
proportional to the number $N_A$ of participant 
nucleons. 
As the energy and $N_A$ increase the role of sea quarks and gluons
increases, they interact and produce, again via strings, particles, and
the number of collisions $\nu$ becomes the relevant 
parameter.
%
%
One should notice that 
multiple inelastic scattering may occur
either internally within a given
nucleon-nucleon collision or externally involving interactions with
different nucleons.

Following Ref.~\citelow{Armesto:SFM}, and taking into account the
above basic 
properties,
we now 
write an expression for the particle pseudo-rapidity density,
\begin{equation}
   \left.\frac{dN}{dy}\right|_{N_AN_A} = 
         N_A \left[ 2 + 2(k-1)\alpha\right] h
                          + (\nuNA - N_A)  2k\alpha h ,   
      \label{eq:Armesto}
\end{equation}
where $h$ is the height of the valence-valence rapidity plateau,
$\alpha$ is the relative weight of the sea-sea (including gluons)
plateau and $k$ is the average number of string pairs per collision.
Elementary multi-scattering arguments\cite{Armesto:SFM} give
	$\nuNA = N_A^{4/3}$.
However, as we mentioned above, the diagram 
corresponding to sea-sea scattering
can be iterated with $k\geq 1$ being, in general, a function of energy.
The number of nucleon-nucleon collisions is, of course,
        $N_A + (\nuNA - N_A) = \nuNA$,
and the number of strings is
        $N_s = N_A\left[ 2 +2(k-1)\right] + (\nuNA - N_A) 2k =
                        2k\nuNA$.
The first term on the right-hand side of Eq.~(\ref{eq:Armesto})
is just a sum over
nucleon-nucleon scattering contributions (including internal parton
multiple scattering) and we can thus write
\begin{equation}
		\left.\frac{dN}{dy}\right|_{N_AN_A} = \left.\frac{dN}{dy}\right|_{pp}
				N_A + (\nuNA - N_A) 2k\alpha h  
																\label{eq:nofusion}
\end{equation}
If external multiple scattering is absent, by putting $\nuNA =
N_A$, one obtains the wounded nucleon model limit;
if $k \gg 1$ we obtain the limit in which
multiple scattering dominates.

\begin{figure}
\begin{minipage}[t]{0.48\textwidth}
  \begin{center}
  \mbox{\includegraphics[width=\textwidth]{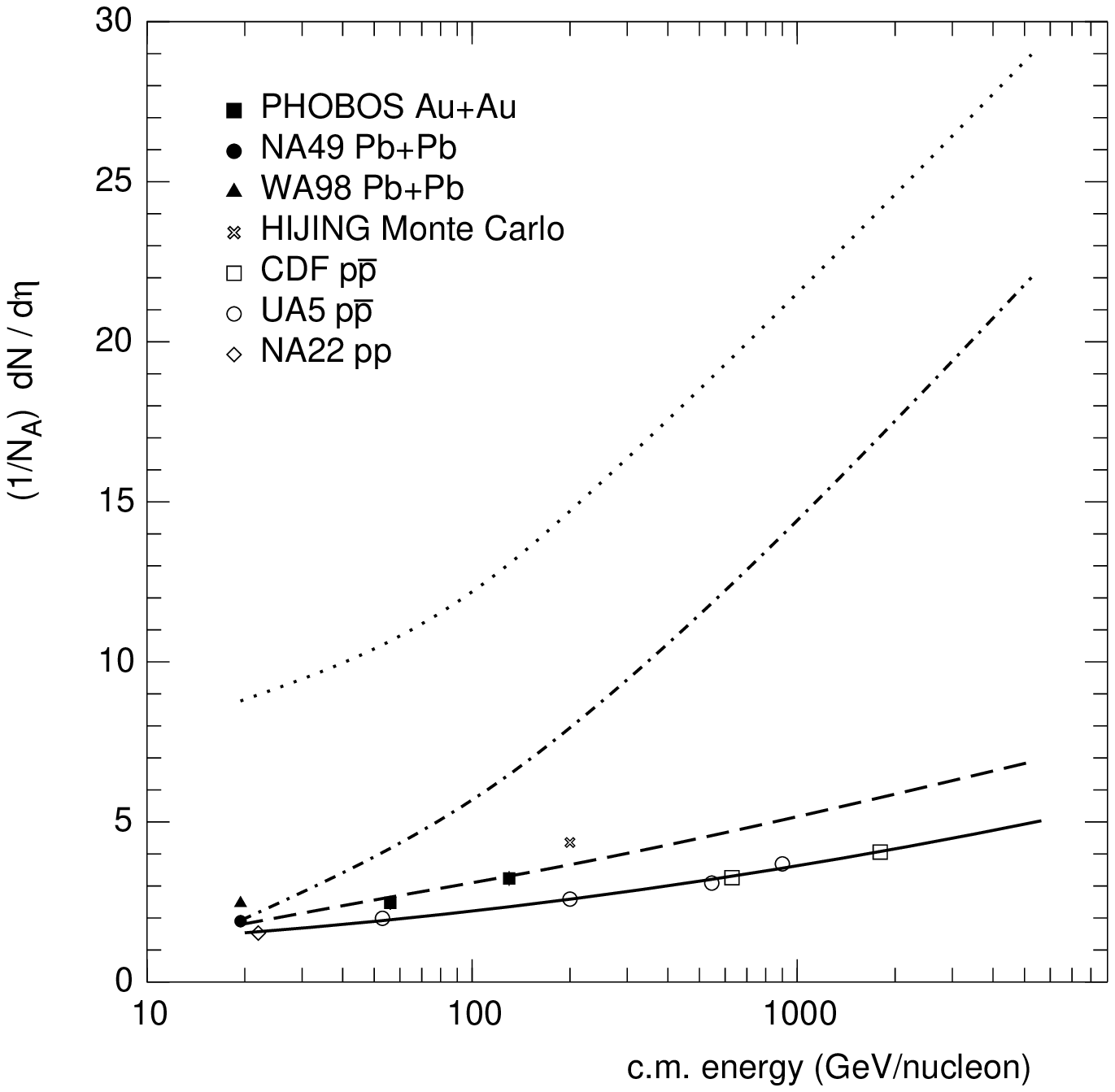}}
  \end{center}
  \caption[Pseudo-rapidity density vs c.m. energy]{Pseudo-rapidity 
	density normalised per participant pair
  as a function of c.m.\ energy. The lines give predictions for the
  wounded nucleon model  (solid line), 
  the pure multicollision approach 
  (dotted line), and the Dual String Model, without fusion 
  Eq.~(\ref{eq:nofusion})
  (dash-dotted line) and with fusion Eq.~(\ref{eq:fusion})
  (dashed line). AA points are taken from Ref.~\citelow{phobos:1,WA98:1,Wang},
	$pp$ and $p\bar p$ from Ref.~\citelow{NA22:b4,UA5:rep,UA5:3,CDF:dNdeta}
  }\label{fig:dNdeta}
\end{minipage}
\hfill
\begin{minipage}[t]{0.48\textwidth}
  \begin{center}
  \mbox{\includegraphics[width=0.9\textwidth]{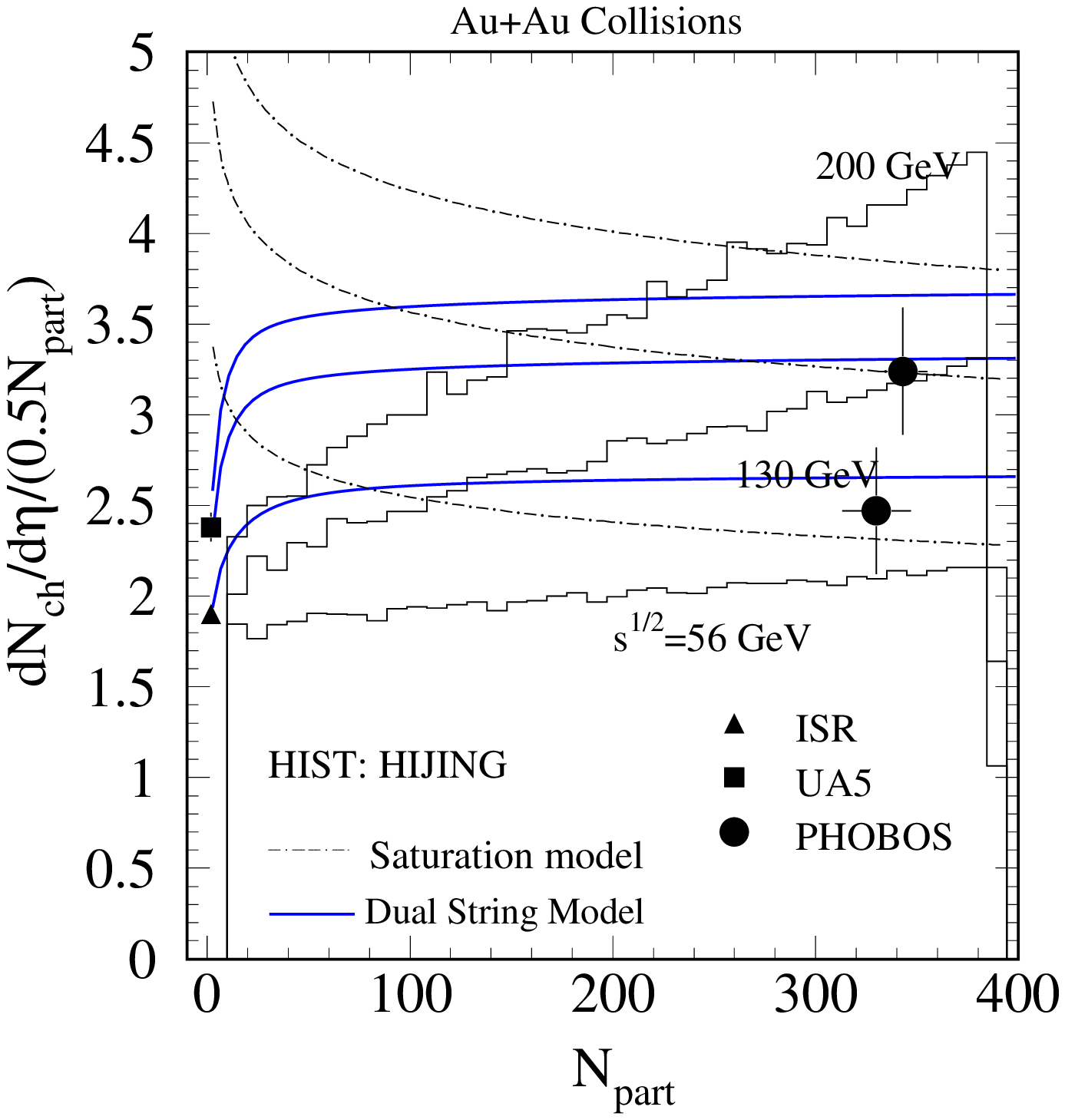}}
  \end{center}
  \caption[Central charged particle rapidity density per 
	participating pair]{Central charged particle rapidity density 
	per participating pair as
  a function of the number of participants. Results of HIJING
  (histograms), EKRT predictions (dot-dashed lines) and DSM predictions
  (solid lines) for central Au+Au collisions at $\sqrt{s}=56,130,200$
  $A$GeV. Also shown are results from $pp$ and $p\bar p$ collisions
  and PHOBOS data (Everything in the figure except the DSM curves is 
  taken from Ref.~\citelow{Wang}.)}\label{fig:wang}
\end{minipage}
  \end{figure}

In Fig.~\ref{fig:dNdeta}, together with the PHOBOS data, 
we have presented the quantity
$\frac{1}{N_A} \left.\frac{dN}{dy}\right|_{N_AN_A}$ as function of the
c.m.\ energy $\sqrt{s}$ for the wounded nucleon model limit---solid
line---and the multiple scattering dominance limit---dotted line.
Assuming that $h$ and $\alpha$ are energy independent (constant
plateaus), the energy dependence of ${dN}/{dy}|_{pp}$,
obtained from a parametrisation of experimental data,\cite{RU:dNdeta}
fixes the energy
dependence of $k$. We find $\alpha = 0.05$ and $h = 0.75$.

In the DSM,
strings may interact by fusing\cite{Pajares:perc+NardiSatz,jotapsi} in the
transverse plane of interaction 
thus modifying the number and the distributions of produced particles:
in particular, due to the vector nature of the colour charge, a
cluster of $m$ strings will emit fewer particles that $m$ separate
strings.\cite{Biro:randomsum} 

The number of strings coming from nucleon multiple scattering ---the
second term in Eq.\ (\ref{eq:Armesto})---is $N_A (N_A^{1/3}-1) 2 k$ and
they occupy the transverse interaction area $\SNA$, which, for central
collisions, is approximately given by
		$\SNA \simeq \pi \left(1.14 N_A^{1/3}\right)^2$   ,	
such that the dimensionless transverse density parameter $\eta$ is
\begin{equation}
	\eta = \left( \frac{r_s}{1.14} \right)^2 2 k N_A^{1/3} 
					(N_A^{1/3} - 1) ,														\label{eq:13}
\end{equation}
where $r_s \simeq 0.2$ fm is the string transverse section
radius. Note that $\eta$ increases with $N_A$ and, via $k$,
also with $\sqrt{s}$.

When fusion occurs, 
Eq.\ (\ref{eq:nofusion}) becomes\cite{RU:dNdeta}
\begin{equation}
	\left.\frac{1}{N_A}\frac{dN}{dy}\right|_{N_AN_A} =
   \left.\frac{dN}{dy}\right|_{pp}  + F(\eta) (N_A^{1/3} - 1) 2 k
	\alpha h	,																			\label{eq:fusion}
\end{equation}
where $F(\eta)$ is the particle production reduction factor,\cite{Braun:Feta}
\begin{equation}
	F(\eta) \simeq \sqrt{ \frac{1-e^{-\eta}}{\eta} }	.		\label{eq:15}
\end{equation}

\begin{figure}
  \begin{center}
  \mbox{\includegraphics[width=0.8\textwidth,height=5cm]{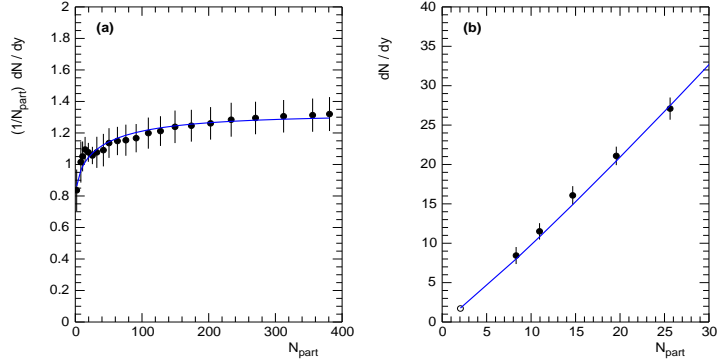}}
  \end{center}
  \caption[Charged particle density]{\textbf{(a)} Charged particle density per 
	participant nucleon versus the number of participants;
	\textbf{(b)} absolute charged particle density
	versus the number of participants. The data from WA98\cite{WA98:1}
	refer to $158 A$ GeV Pb+Pb collisions (filled circles), the open
	circle refers to $pp$ collisions\cite{DeMarzo}; the solid line
	results from Eq.\ (\ref{eq:fusion}) with $\alpha=0.11$ and
  $h=0.77$.}\label{fig:wa98}
  \end{figure}

It can now easily be shown\cite{RU:dNdeta:2} that
the DSM with fusion predicts saturation of the particle rapidity
densities per participant pair of nucleons as $N_A$ increases.
This prediction is compared to other models\cite{Wang,EKRT,EKRT:2}
in figure \ref{fig:wang} and to experimental data from 
WA98\cite{WA98:1} in figure \ref{fig:wa98}.

Furthermore, it is to be noted that
the predictions for particle densities in central Pb+Pb collisions
   of the DSM without fusion and of the DSM with fusion are very
   different at $\sqrt{s} = 200$ AGeV (RHIC) and at $\sqrt{s} = 5.5$
   ATeV (LHC) as can be seen in the following table:
	\begin{center}
  \begin{tabular}{c|c|c}
	\hline
	c.m.\ energy   & 200 AGeV & 5.5 ATeV \\
	\hline
	without fusion & 1500     &  4400    \\
	with fusion    &  700     &  1400    \\
	\hline
  \end{tabular}
	\end{center}
Of course this model is essentially soft. The
   parameters of the elementary collision densities, $h$ and $\alpha$,
   were assumed constant, all the energy dependence being attributed
   to the parameter $k$, the average number of string pairs per
   elementary collision.
	 If $h$ and $\alpha$ are allowed to grow with energy, as a result,
   for instance, of semi-hard effects, the parameter $k$ may then have
   a slower increase than the one obtained here.

Finally, one should consider the idea that string fusion eventually
leads to a situation of percolation\cite{Pajares:perc+NardiSatz,jotapsi}
with the formation of extended regions of
colour freedom, with the features of the expected Quark-Gluon Plasma.
Indeed the parameter	 $\eta$ at the CERN-SPS
   has the value $\eta \approx 1.8$, larger than the critical
   density ($\eta_c \approx 1.12\div1.17$) which means that percolation
   transition is already taking place at $\sqrt{s} = 20$ AGeV, even
   allowing for non-uniform matter distribution in the nucleus
   ($\eta_c\approx 1.5$);\cite{perc:98} this result is valid
	 even with $k=1$.
	 The observed anomalous $J/\psi$ suppression\cite{NA50:QGP} may
   then be a signature of the percolation transition to the
   Quark-Gluon Plasma.\cite{Pajares:perc+NardiSatz,jotapsi} 

Indeed, in our simple approach,\cite{jotapsi}
$J/\psi$ and Drell-Yan production are treated as rare events:
this implies that their ratio is given by the product of two
functions, one describing absorption of $J/\psi$
(which we assume as usual to be exponential in the amount
of matter longitudinally traversed), the other describing
$J/\psi$ ($c\bar c$) suppression due to Debye screening.
If we take the drastic position that the latter is
100\% effective if there is
percolation, and ineffective otherwise,
then screening is described by the probability of non-percolation,
which can be parametrised as
\begin{equation}
	P_{\text{non-perc}}(\eta) = \left[
					1 + \exp\left( \frac{\eta-\eta_c}{a_c} \right)
				\right]^{-1}  ,
\end{equation}
with $a_c$ a parameter linked to the finite size of the nuclear
system. 
Thus we see that the onset of the phase transition is characterised by
a change in the curvature of the $J/\psi$ over D.Y.\ ratio from positive
(during absorption) to negative.
This however is only a qualitative description: a quantitative one
should probably take into account more details of the process (e.g.,
geometry varying with impact parameter, resonances, $\dots$).

In conclusion,
the DSM is a model with two components, the valence-valence component
and the sea-sea component, the sea-sea component increasing its
importance with energy and number of participants.
This is somewhat similar to the HIJING Monte Carlo model, with soft
and hard components.
On the other hand, with fusion the DSM behaves, for large $N_A$,
similarly to the EKRT model, but with strict saturation of the
particle density per participant nucleon.
However, in the original EKRT model the saturation criterion in the
transverse plane is stronger than in case of fusion of strings.
Here, saturation in the interaction area is asymptotic (when
$\eta\to\infty$) while in the EKRT model it occurs
at finite density.
This causes the decrease of the particle density 
with $N_A$ in the EKRT original
model.

Probably different explanations, such as the ones based on string
fusion, parton saturation, parton shadowing, are in some sense dual
and refer to the same underlying physics.\cite{Armesto:SFM}
What is becoming clear is that saturation of particle density puts
strong constraints in models, and limits the rise of the
(pseudo-)rapidity plateau at RHIC and LHC.

\section*{Acknowledgements}
R.U. gratefully acknowledges the financial support of the
Fundação Ciência e Tecnologia via the ``Sub-Programa Ci\^encia 
e Tecnologia do $2^o$ Quadro Comunit\'ario de Apoio.''

\bibliographystyle{prstyR}
\bibliography{abbrevs,bibliography}

\end{document}